\documentstyle[12pt,psfig,a4]{article} 
\begin{document} 

\begin{titlepage}

\hfill TUM/T39--98--17

\begin{center}
{\Large\bf Nucleon Polarizibilities for Virtual Photons}\footnote{Work 
supported in part by DFG and BMBF.}

\bigskip

\bigskip

{\large J. Edelmann, N. Kaiser, G. Piller and  W.Weise}

\bigskip

Physik-Department, Technische Universit\"{a}t M\"{u}nchen, \\
D-85747 Garching, Germany 

\end{center}

\vskip 2cm

\begin{abstract} 
We generalize the sum rules for the nucleon electric plus magnetic
polarizability $\Sigma=\alpha+\beta$ and for the nucleon spin-polarizability 
$\gamma$, to virtual photons with $Q^2>0$. The dominant low energy cross
sections are represented in our calculation by one-pion-loop graphs of 
relativistic baryon chiral perturbation theory and the $\Delta(1232)$-resonance
excitation. For the proton we find good agreement of the calculated
$\Sigma_p(Q^2)$ with empirical values obtained from integrating up
electroproduction data for $Q^2<0.4$ GeV$^2$. The proton spin-polarizability 
$\gamma_p(Q^2)$ switches  sign around $Q^2= 0.4$ GeV$^2$ and it joins smoothly 
the "partonic" curve, extracted from polarized deep-inelastic scattering, 
around $Q^2=0.7$ GeV$^2$. For the neutron our predictions of $\Sigma_n(Q^2)$ 
and $\gamma_n(Q^2)$ agree reasonably well at $Q^2=0$ with existing 
determinations. Upcoming (polarized) electroproduction experiments will be able
to test the generalized polarizability sum rules investigated here.

\end{abstract}
\end{titlepage}

\section{Introduction}
The electromagnetic interaction in the form of real photon absorption and
(inelastic) electron scattering is one of the major experimental tools to
study the excitation spectrum of the nucleon. It consists of three main 
components: non-resonant (multi-) meson production near the respective
thresholds, excitation of baryon resonances ($N^*$,  $\Delta$ etc.) with
definite quantum numbers and the parton (quark and gluon) distributions as 
revealed in deep inelastic lepton scattering. 

Unitarity (the optical theorem) and forward dispersion relations connect 
integrals over the whole nucleon excitation spectrum to certain low energy
parameters. Prominent examples are Baldin's sum rule \cite{baldin} and the 
Gerasimov-Drell-Hearn sum rule \cite{gdh}. In the first case the sum of the 
nucleon electric and magnetic polarizability $\Sigma=\alpha+\beta$ is equated 
to the integral over the real photon absorption cross section $\sigma_{\rm 
tot}(\omega)$ weighted by the inverse squared photon lab energy $\omega$,
\begin{equation} \Sigma=\alpha+\beta = {1\over 2\pi^2}
\int_{\omega_{th}}^\infty {d\omega \over \omega^2} \sigma_{tot}(\omega)\,\,.
\end{equation} 
Here, $\omega_{th}=m_\pi(1+m_\pi/2M)=150$ MeV is the pion photoproduction
threshold, and $m_\pi$ and $M$ denote the pion and nucleon mass, respectively.
The resulting value for the proton $\alpha_p+\beta_p= (14.2\pm 0.3) \cdot 
10^{-4}$ fm$^3$ \cite{damash} (recently reevaluated \cite{babusci} with 
improved photoproduction data to be $\alpha_p+\beta_p= (13.69\pm 0.14) \cdot 
10^{-4}$ fm$^3$) is often used in the analysis of low energy proton Compton
scattering in order to facilitate the difficult separation of the electric
polarizability $\alpha_p$ and magnetic polarizability $\beta_p$ of the proton. 
On the theoretical side chiral perturbation theory permits systematic
calculations of this quantity \cite{pola,chirstruc,pola2}. At leading order in 
the quark mass (or, equivalently, the pion mass) expansion one finds $\alpha_p+
\beta_p= 11 e^2 g_{\pi N}^2/(768 \pi^2 M^2 m_\pi)= 15 \cdot 10^{-4}$ fm$^3$, a 
number which is in almost perfect agreement with the  empirical determination. 
However this result does not include the strong paramagnetic effects from the 
$\Delta(1232)$-resonance. A next-to-leading order calculation leads to 
$\alpha_p + \beta_p = (14 \pm 4) \cdot 10^{-4}$ fm$^3$ \cite{pola2}. At this 
order one can indeed establish (for the proton) the important cancelation  
between diamagnetic pion-loop effects ($\sim \ln m_\pi$) and paramagnetic 
effects from the $\Delta(1232)$-resonance. The theoretical uncertainty ($\pm
4$) results from some not very well known low energy parameters entering the
calculation and from estimates of the $\Delta(1232)$-contribution itself.  

The Gerasimov-Drell-Hearn sum rule \cite{gdh} connects the squared (proton or
neutron) anomalous magnetic moment $\kappa$ to an integral over the difference
of the helicity $1/2$ and $3/2$ photoabsorption cross sections $\sigma_{1/2,
3/2}(\omega)$ weighted by the inverse photon lab energy $\omega$, 
\begin{equation} -{\pi e^2 \kappa^2 \over 2M^2} = \int_{\omega_{th}}^\infty 
{d\omega \over \omega} \Big[\sigma_{1/2}(\omega)-\sigma_{3/2}(\omega) \Big]
\,\,. \end{equation} 
At present no direct measurements of these helicity cross sections exist. They
have been reconstructed from the single pion photoproduction 
multipoles \cite{karliner,sandorfi} and from estimates of the two-pion 
photoproduction contribution. With this input one finds a qualitative
agreement at the 20--30\% level between the left and right hand side of eq.(2),
but there remains a notorious mismatch in sign once one considers the 
proton-neutron difference. Upcoming experiments at Mainz (MAMI) and Bonn (ELSA)
are devoted to measuring precisely these helicity photoabsorption cross 
sections $\sigma_{1/2,3/2}(\omega)$ and these measurements will help to clear
up the situation concerning the proton-neutron difference and the validity of
the Gerasimov-Drell-Hearn sum rule itself. 

From a purely theoretical point of view the Drell-Hearn-Gerasimov sum rule
stands or falls with the validity of the no-subtraction hypothesis, since the
other ingredients in its derivation, the optical theorem and the Compton low
energy theorem, are beyond any doubt. Stated differently the
Drell-Hearn-Gerasimov sum rule hinges essentially on the high energy
($\omega\to \infty$) behavior of the (spin-flip) forward Compton amplitude. 

On much safer ground is the sum rule for the so-called spin-polarizability
$\gamma$ since it does not require the no-subtraction hypothesis. The nucleon 
spin polarizability $\gamma$ is the integral over the difference of the 
helicity $1/2$ and $3/2$ photoabsorption cross sections weighted by the 
inverse  third power of the photon lab energy $\omega$, 
\begin{equation} \gamma = {1\over 4\pi^2} \int_{\omega_{th}}^\infty {d\omega
\over \omega^3} \Big[\sigma_{1/2}(\omega)-\sigma_{3/2}(\omega)\Big]\,\,. 
\end{equation} 
Semi-empirical determinations using the pion photoproduction multipoles give 
$\gamma_p = -1.34\cdot 10^{-4}$ fm$^4$ and $\gamma_n = -0.38 \cdot 10^{-4}$ 
fm$^4$
\cite{sandorfi}. The calculation presented in ref.\cite{chirstruc} allows to
understand these values in terms of compensative effects from relativistic 
pion-loops and the $\Delta(1232)$-resonance, $\gamma_p = (2.16-3.66 )\cdot
10^{-4}$ fm$^4=-1.50 \cdot 10^{-4}$ fm$^4$,  and $\gamma_n = (3.20-3.66) \cdot
10^{-4}$ fm$^4=-0.46 \cdot 10^{-4}$ fm$^4$. Obviously the $\Delta(1232)$-effect
is dominant in both cases, but the relativistic pion-loop effects are also
important in order to reproduce the difference between the proton and neutron
value of $\gamma$. It should also be mentioned that the calculation of 
$\gamma_{p,n}$ in ref.\cite{chirstruc} is not a complete and systematic 
order-by-order calculation, but it seems rather convincing that the 
relativistic pion-loops and the $\Delta(1232)$-resonance account well for the
low energy dynamics encoded in $\gamma_{p,n}$ as well as in $\Sigma_{p,n}$.  

The purpose of this work is to generalize the Baldin sum rule for
$\Sigma=\alpha+\beta$ and the sum rule for $\gamma$ to virtual photons with
$Q^2>0$ in the form of energy-weighted integrals over measurable
electroproduction cross sections. We will present theoretical predictions
for $\Sigma_{p,n}(Q^2)$ and $\gamma_{p,n}(Q^2)$ by evaluating the contributions
from the relativistic pion-loop and the $\Delta(1232)$-resonance as done in
ref.\cite{chirstruc,BKM}. In the case of $\Sigma_p(Q^2)$ we can compare with an
(approximate) empirical determination using as input electroproduction data 
from DESY \cite{data}. In the high-$Q^2$ regime the generalized polarizability 
sum rules are given by integrals over the (partonic) structure functions 
$F_1(x)$ and $g_1(x)$ times a characteristic power of $1/Q^2$. We investigate 
furthermore whether the low energy hadronic side (given in terms pion-loops 
and the $\Delta(1232)$-resonance) extrapolated upward and the partonic side 
extrapolated downward match at some intermediate scale. We find that this is 
indeed the case at $Q^2 \approx 0.5 \dots 0.7$ GeV$^2$. This  
means that the quantities $\Sigma_{p,n}(Q^2)$ and $\gamma_{p,n}(Q^2)$ are not 
very sensitive to the details of the nucleon excitation spectrum. Our 
predictions for $\Sigma_{p,n}(Q^2)$ and $\gamma_{p,n}(Q^2)$ can
be tested once the relevant electroproduction cross section are available. 
There exist extended experimental programs to perform such measurements at the
electron accelerator laboratories MAMI (Mainz), ELSA (Bonn) and TJNAF (Newport
News).   

Our paper is organized as follows. In section 2 we present the necessary
formalism, the transverse virtual forward Compton amplitude, the optical
theorem and the definition of the generalized polarizability sum rules for
$\Sigma(Q^2)$ and $\gamma(Q^2)$ in terms of electroproduction cross
sections. In section 3 we discuss our results for the proton and
the neutron. Section 4 ends with a summary. The appendix includes explicit 
formulas for the respective virtual forward Compton amplitudes as they derive
from the relativistic one-pion loop  graphs of chiral perturbation and
$\Delta(1232)$-excitation tree diagrams.    

\section{Formalism}
In this section we present the formalism for transverse virtual forward Compton
scattering off the nucleon. The more familiar case of real photon forward
Compton scattering can be easily recovered by simply setting $Q^2=0$ and will 
therefore not be treated here separately. 
  
We consider virtual photons with $-q^2=Q^2>0$. The forward transverse amplitude
for the process  $\gamma^*(q,\epsilon) + N(p) \to \gamma^*(q,\epsilon\,') +
N(p)$  reads 
\begin{equation} T = f_1(\omega^2,Q^2) \, \vec \epsilon\,'^*\cdot \vec\epsilon 
+ f_2(\omega^2,Q^2) \, i \omega\, \vec \sigma \cdot (\vec \epsilon\,'^*\times
\vec  \epsilon\,)\,\,,  \end{equation}
with $f_{1,2}(\omega^2,Q^2)$ the generalizations of the spin non-flip and
spin-flip forward Compton amplitudes, respectively. Crossing symmetry implies 
that these are even functions of $\omega$. The polarization vectors $\epsilon$
and $\epsilon\,'$ satisfy the gauge and transversality conditions $\epsilon
\cdot p =\epsilon \,'\cdot p =\epsilon\cdot q =\epsilon\,' \cdot q =0$. The
variable $\omega$ is the virtual photon energy in the nucleon rest frame, $
\omega = p\cdot q/M$. In terms of the Lorentz-invariant Mandelstam variable
$s=(p+q)^2$ one has $\omega=(s-M^2+Q^2)/2M$.

The Born contributions to $f_{1,2}(\omega^2,Q^2)$ come from the direct and
crossed nucleon pole diagram with only a nucleon in the intermediate state. 
These contributions can be expressed in terms of the (on-shell) electric and 
magnetic nucleon form factors. At the real photon point $Q^2=0$ the Born terms
in the limit $\omega\to 0$ agree exactly with the Compton low energy theorems, 
$f_1(0,0)=-e^2Z^2/(4\pi M), \, f_2(0,0) = -e^2 \kappa^2/(8\pi M^2)$.  Here $Z,
\, \kappa$ and $M$ denote the electric charge, anomalous magnetic moment and
mass of the nucleon (proton or neutron).  

We consider here only non-Born contributions to $f_{1,2}(\omega,Q^2)$ which 
correspond the inelastic electron-nucleon processes.  Due to
crossing symmetry one can write down the following representation,
\begin{eqnarray} f_1(\omega^2, Q^2) &=& {e^2 \over 4\pi M} \Big[ A_1(s, Q^2)
+A_1(2M^2-2Q^2-s, Q^2) \Big] \,\,, \\ \omega \, f_2(\omega^2, Q^2) &=& {e^2 
\over 4\pi M} \Big[ A_2(s, Q^2) -A_2(2M^2-2Q^2-s, Q^2)\Big]\,\,.\end{eqnarray} 
The dimensionless functions $A_{1,2}(s,Q^2)$ have only a
right hand cut starting at the single pion electroproduction threshold $s_{th}=
(M+m_\pi)^2$ \cite{BKM}. In other words, the construction is made such that 
$A_{1,2}(s,Q^2)$ receive contributions only from all direct 
(virtual) Compton diagrams and the contributions from all crossed diagrams are
generated by the (anti)-symmetrization procedure $s\to 2M^2-2Q^2-s$. 

In the physical region of inelastic electron-nucleon scattering
($s>(M+m_\pi)^2$) the optical theorem relates the imaginary parts of $f_{1,2}
(\omega^2, Q^2)$ to nucleon structure functions or, equivalently, to inclusive 
electroproduction cross sections,
\begin{eqnarray} {\rm Im}\,f_1(\omega^2, Q^2) &=& {e^2 \over  4M}
W_1(\omega, Q^2) = {\omega \over 4\pi } \bigg( 1 -{Q^2 \over 2M\omega} \bigg)
\sigma_T(\omega ,Q^2) \,\,, \\ {\rm Im}\,f_2(\omega^2, Q^2) &=& {e^2 
\over 4 M^2}  \bigg[ G_1(\omega, Q^2) -{Q^2 \over M\omega} G_2(\omega, Q^2)
\bigg] \nonumber\\ &=& {1\over 8\pi} \bigg( 1-{Q^2 \over 2M\omega} \bigg) \Big[
\sigma_{1/2}(\omega, Q^2) - \sigma_{3/2}(\omega, Q^2) \Big] \,\,. 
\end{eqnarray} 
Here $W_1(\omega, Q^2)$ and $G_{1,2}(\omega, Q^2)$ are the usual unpolarized
and polarized nucleon structure functions, respectively. $\sigma_T(\omega, 
Q^2)$ is the transverse electroproduction cross section measured in unpolarized
inelastic electron-nucleon scattering. Furthermore, $\sigma_{1/2,3/2}(\omega,
Q^2)$ are  the helicity cross sections measured with polarized leptons on
polarized nucleons with parallel spins pointing either in opposite or in the 
same direction. 

As mentioned in the introduction the combination of a (once-subtracted) 
forward dispersion relation and the optical theorem leads to the sum rules for
the sum of electric and magnetic polarizability $ \Sigma = \alpha + \beta$ 
(Baldin's sum rule), and for the so-called spin-polarizability $\gamma$. We now
generalize these sum rules, eqs.(1,3), to virtual photons with $Q^2>0$ in the 
(most natural) form,
\begin{eqnarray} \Sigma(Q^2) &=&  {1\over 2\pi^2} \int_{\omega_{th}}^\infty {d
\omega \over \omega^2}\,\sigma_T(\omega ,Q^2) \,\,, \\ \gamma(Q^2) &=&  {1\over
4 \pi^2} \int_{\omega_{th}}^\infty {d \omega \over\omega^3}\,\Big[ \sigma_{1/2}
(\omega,Q^2)- \sigma_{3/2}(\omega, Q^2) \Big]  \,\,, \end{eqnarray}
with $\omega_{th} = m_\pi +(m_\pi^2+Q^2)/2M$ the (single) pion 
electroproduction threshold. Note that our definition of the generalized
polarizabilies $\Sigma(Q^2)$ and $\gamma(Q^2)$ is different from the one in 
ref.\cite{hemmert}. In this work (non-forward) Compton scattering with a 
virtual photon in the initial state and a real photon in the final state was
considered.   

Assuming that the functions $A_{1,2}(s,Q^2)$ introduced in 
eqs.(5,6) satisfy a (once-subtracted) dispersion relation, which is indeed the
case for the relativistic pion-loop diagrams, we arrive after some algebraic
manipulation at the following representation of the generalized
polarizabilities, 
\begin{eqnarray} \Sigma(Q^2) &=&  {2e^2 M \over \pi Q^4} \Big[ A_1(M^2,Q^2)
-A_1(M^2-Q^2,Q^2)-Q^2 A_1'(M^2-Q^2,Q^2) \Big] \,\,, \\ \gamma(Q^2) &=&  
{ 4e^2 M^2 \over \pi Q^6} \bigg[ A_2(M^2,Q^2)-A_2(M^2-Q^2,Q^2)-Q^2 A_2'(M^2-
Q^2,Q^2)\nonumber \\ & & \qquad\quad -{Q^4\over2} A_2''(M^2-Q^2,Q^2)\bigg]\,\,.
\end{eqnarray}
Here the prime denotes the partial derivative with respect to the variable $s$.
In order to arrive at the representations eqs.(11,12) it is important that 
$A_{1,2}(s,Q^2)$ have only right hand cuts, with their discontinuities 
proportional to the electroproduction cross sections in eqs.(9,10).  
To make theoretical predictions for $\Sigma(Q^2)$ and $\gamma(Q^2)$ we will 
compute $A_{1,2}(s,Q^2)$ 
for the proton and the neutron by evaluating all 52 one-pion loop diagrams of 
relativistic baryon chiral perturbation theory and the relativistic 
$\Delta(1232)$-excitation tree-graph (employing the Rarita-Schwinger
formalism). These are the dominant processes at low energies where the 
energy-weighted sum rules eqs.(9,10) for $\Sigma(Q^2)$ and $\gamma(Q^2)$ are
almost saturated. The explicit formulas for the various
contributions to $A_{1,2}(s,Q^2)$ are given in the appendix.  

At large energies and momentum transfer the nucleon structure functions show
a scaling behavior, i.e. they depend only on the dimensionless Bjorken variable
$x= Q^2/(2M\omega)$. For such kinematics one has $W_1(\omega,Q^2) = F_1(x)$ and
$G_1(\omega, Q^2)- (Q^2/M\omega) G_2(\omega,Q^2) = (M/\omega) g_1(x)$, with
the unpolarized and polarized nucleon structure functions $F_1(x)$ and $g_1(x)$
in the scaling limit. The contribution of the second spin-dependent structure
function $g_2(x)$ is suppressed by a prefactor $-(2x M/Q)^2$. As a 
consequence of the scaling relations one finds that the large-$Q^2$ behavior 
of the generalized polarizabilities $\Sigma(Q^2)$ and $\gamma(Q^2)$ is given by
certain integrals over the nucleon structure functions $F_1(x)$ and $g_1(x)$
in the scaling limit times a power of $1/Q^2$, 
\begin{eqnarray} \Sigma(Q^2) &=& {2e^2 M\over \pi Q^4} \int_0^1 dx {x\over 1-x}
F_1(x) \,\,, \\ \gamma(Q^2) &=& {4e^2 M^2\over \pi Q^6} \int_0^1 dx {x^2\over 
1-x} g_1(x) \,\,. \end{eqnarray}
The relevant integrals on the right hand side can be evaluated with existing
parametrizations \cite{parton1,parton2} of the structure functions $F_1(x)$ and
$g_1(x)$. It is interesting to see whether the predictions of our 
approach valid for small $Q^2$, which explicitly treats relativistic pion-loops
and the $\Delta(1232)$-excitation, matches the QCD-asymptotics valid for large
$Q^2$ at some intermediate scale $Q^2$.   

\section{Results and discussion}
In this section we present and discuss our results for $\Sigma_{p,n}(Q^2)$
and $\gamma_{p,n}(Q^2)$. The formulas for relativistic pion-loop contributions
(given in the appendix) depend on the $\pi N$-coupling constant $g_{\pi N}$ 
for which we use $g_{\pi N}=13.4$. The evaluation of the relativistic 
$\Delta$-excitation graphs requires more information. First there
is the $\Delta \to N \gamma$ transition strength $\kappa^*$ for which we use
the large-$N_c$ relation to the isovector nucleon magnetic moment, $\kappa^*=
3(1+\kappa_p-\kappa_n)/2\sqrt2= 5.0$. Second, there is the Rarita-Schwinger 
off-shell parameter $Y$ with the empirical band $-0.8 < Y <1.7$ \cite{benmer}, for which we choose $Y=0$. This value is essentially fixed from the
$\Delta(1232)$-contribution to $\Sigma(0)=\alpha+\beta$ (see the discussion
below). Third, the $\Delta\to N\gamma^*$-transition occurs at non-zero $Q^2$
and thus there is also a transition form factor $G_\Delta(Q^2)$. We use 
\begin{equation} G_\Delta(Q^2 ) = { \exp(-0.2 Q^2/ {\rm GeV}^2) \over (1+1.41
Q^2 / {\rm GeV}^2)^2} \end{equation}
as extracted from pion electroproduction in the $\Delta(1232)$-resonance
region in ref.\cite{dform}. Apart from the exponential function in the 
numerator, $G_\Delta(Q^2)$ is the usual dipole fit to the proton electric form
factor. 

We start the discussion with the values of $\Sigma_{p,n}(Q^2)$ and 
$\gamma_{p,n}(Q^2)$ at the real photon point $Q^2=0$. With the parameter input 
mentioned before we get
\begin{equation} \Sigma_p(0) = (5.48 + 8.23)\cdot 10^{-4}\,{\rm fm}^3 = 13.71
\cdot 10^{-4}\,{\rm fm}^3 \,\,, \end{equation}
\begin{equation} \Sigma_n(0) = (8.90 + 8.23)\cdot 10^{-4}\,{\rm fm}^3 = 17.13
\cdot 10^{-4}\,{\rm fm}^3 \,\,,\end{equation}
where the first number in the bracket comes from the pion-loops and the second
one from the $\Delta$-excitation. The latter is given by the expression
\begin{equation} \Sigma_{p,n}^{(\Delta)}(0) = {e^2 \kappa_*^2 \over 18 \pi M
M_\Delta^2} \bigg[ {M_\Delta^2 +M^2 \over M_\Delta^2 -M^2} - 4Y(1+2Y) \bigg] 
\,\,, \end{equation}
and its theoretical value $8.23 \cdot 10^{-4}$ fm$^3$ for $Y=0$ is about 20\% 
larger than the quasi-empirical value $7 \cdot 10^{-4}$ fm$^3$ obtained in
ref.\cite{zang}. Note that the sum of relativistic pion-loops and the
$\Delta$-excitation with  $Y=0$ reproduces very well the empirical value for
the proton, $\Sigma_p(0) = (13.69\pm 0.14)\cdot 10^{-4}$ fm$^3$ \cite{babusci}.
However, the same diagrams overestimate the empirical value for the neutron, 
$\Sigma_n(0)= (14.40\pm 0.66) \cdot 10^{-4}$ fm$^{3}$ \cite{babusci} by about
20\%. A similar problem was also encountered in the heavy baryon chiral
perturbation theory calculation of ref.\cite{pola2}. It originates from the
fact that due to certain numerical coefficients \cite{pola2} the diamagnetic
effects from the pion-loops at next-to-leading order ($\sim \ln m_\pi$) are 
less strong for the neutron than for the proton.     

There is no need to reiterate the values for $\gamma_{p,n}(0)$ given in the 
introduction and in ref.\cite{chirstruc}, since $\gamma_{p,n}(0)$ does not
depend on the off-shell parameter $Y$. This feature holds even for the 
full $Q^2$-dependent $\gamma_{p,n}(Q^2)$. The reason is that the 
$Y$-dependent term in  $A_{2,p,n}^{(\Delta)}(s,Q^2)$ (see appendix) is just a 
quadratic polynomial in $s$, which makes zero contribution to $\gamma_{p,n}
(Q^2)$ according to eq.(12). This independence of the off-shell parameter $Y$
is of course a very welcome feature.

Next, we come to the $Q^2$-dependent quantities $\Sigma_{p,n}(Q^2)$ as 
calculated in our approach. They are shown in Fig.1 for the proton and the
neutron in the region $0 < Q^2<0.4$ GeV$^2$. The dashed and dotted lines give
the relativistic pion-loop and $\Delta(1232)$-contribution, respectively. The
full lines correspond to the sum of both contributions. In Fig.2 we show the
result of  our calculation (full line) together with an evaluation of the sum
rule eq.(9) using as input for $\sigma_T(\omega, Q^2)$ the parametrized
electroproduction data of ref.\cite{data} (dashed line). Since no 
transverse-longitudinal
separation was done for these electroproduction data one should assign at least
a $\pm 15\%$ error band to the dashed line in Fig.2. Note also that the dashed
curve obtained from integrating the unpolarized electroproduction data of 
ref.\cite{data} does not exactly extrapolate the value $\Sigma_p(0) = 13.7\cdot
10^{-4}$ fm$^3$ at the real photon point $Q^2 =0$. Within the uncertainty of 
the empirical $\Sigma_p(Q^2)$ (obtained from integrating the electroproduction
data) there is good agreement with our calculation. Note that  the
$Q^2$-dependence of the pion-loops is entirely due to the pion and nucleon
propagators in the loop-diagrams, and the $Q^2$-dependence of the 
$\Delta(1232)$-contribution is  essentially due to the phenomenological 
$\Delta\to N \gamma^*$ transition form factor $G_\Delta(Q^2)$ given in eq.(15).
In Fig.3 we show the results for $\Sigma_{p,n}(Q^2)$ of our calculation (full 
lines) together with the partonic prediction eq.(13) extrapolated downward to 
$Q^2=0.5$ GeV$^2$. The integrals $\int_0^1 dx\, 2xF_1(x)/(1-x)$ were evaluated 
with parametrized parton distributions of ref.\cite{parton1} taken at $Q^2=1$ 
GeV$^2$. The resulting values of these integrals are $0.453$  for the proton 
and $0.313$ for the neutron. It is interesting to observe that the low 
$Q^2$-behavior given by the pion-loops and the $\Delta(1232)$-resonance matches
approximately  with the downward extrapolated partonic curve at $Q^2=0.5$
GeV$^2$. The integrated quantities $\Sigma_{p,n}(Q^2)$ are obviously not
very sensitive to details of the nucleon excitation spectrum. 

Fig.4 shows the $Q^2$-dependent spin-polarizabilities $\gamma_{p,n}(Q^2)$ for
the proton and the neutron in the region $0< Q^2 < 0.6$ GeV$^2$. For these
quantities a strong cancelation between (positive) pion-loop contributions and
(negative) $\Delta(1232)$-contributions is taking place. For very small $Q^2$ 
the $\Delta(1232)$-effects are actually dominant as witnessed by the negative
values at the real photon point $Q^2=0$. Note that the 
$\Delta(1232)$-contribution decreases faster in magnitude (due to the
transition form factor $G_\Delta(Q^2)$) than the pion-loop contribution. As a
consequence $\gamma_{p,n}(Q^2)$ pass through zero around $Q^2\simeq 0.4$ 
GeV$^2$. At present there are not enough data for the helicity cross sections
$\sigma_{1/2,3/2}(\omega, Q^2)$ such that one could integrate them up to get an
empirical determination of $\gamma_{p,n}(Q^2)$. However these will be measured
in the near future at the electron accelerator laboratories MAMI, ELSA and
TJNAF. In particular one can then test our result for the transition through 
zero of $\gamma_{p,n}(Q^2)$ at $Q^2 \simeq 0.4$ GeV$^2$. Of course, the
one-loop chiral perturbation theory treatment of the pion-cloud contribution is
likely to become uncertain at $Q^2>0.5$ GeV$^2$. Effects beyond one-loop, such
as form factors of the interacting $\gamma^* \pi N$ system, can make these
contributions softer so that the zero of $\gamma_{p,n}(Q^2)$ may be shifted to
somewhat smaller values of $Q^2$. 

Finally, we show in  Fig.5 the results  of our calculation for $\gamma_{p,n}
(Q^2)$  (full lines) together with the partonic prediction (dashed lines)
extrapolated downward. The integrals $\int_0^1 dx x^2 g_1(x)/(1-x)$ were
evaluated with the parton distributions of ref.\cite{parton2} taken at $Q^2=1$
GeV$^2$. The resulting values for these integrals  are $2.72\cdot 10^{-2}$ for
the proton and $-1.52\cdot 10^{-3}$ for the neutron. Again one observes a
smooth transition from the hadronic side (pion-loops plus $\Delta(1232)$) to
the partonic side around $Q^2 = (0.6-0.7)$ GeV$^2$ in the case of the proton. 
The partonic prediction for $\gamma_n(Q^2)$ is very small and negative,
whereas the sum of pion-loop and $\Delta(1232)$-contributions is positive in
the low-$Q^2$ region. There is an almost complete cancelation of 
the pion-loop and $\Delta(1232)$-contributions at $Q^2=0.5$ GeV$^2$, and this
zero would move upward with a slight change of the transition form 
factor $G_\Delta(Q^2)$. Within the accuracy of our approach one can say that
there is no mismatch for the neutron's $\gamma_n(Q^2)$ between the 
hadronic  prediction at low $Q^2$ and the partonic prediction extrapolated 
downward.

\section{Summary}     
In this work we have generalized the sum rules for the nucleon 
electric plus magnetic polarizabilities, $\Sigma = \alpha+\beta$, and the 
so-called spin-polarizability $\gamma$ to virtual photons with $Q^2>0$. Once 
the respective unpolarized and polarized nucleon electroproduction cross
sections $\sigma_T(\omega, Q^2)$ and $\sigma_{1/2,3/2}(\omega, Q^2)$ are
measured these sum rules can be evaluated empirically. We have presented a 
calculation of the quantities $\Sigma_{p,n}(Q^2)$ and $\gamma_{p,n}(Q^2)$ at 
low $Q^2$ in terms of relativistic pion-loop graphs of chiral perturbation
theory and the $\Delta(1232)$-excitation. The off-shell parameter of the 
$\Delta N\gamma$-vertex ($Y=0$) was fixed through the value $\Sigma_p(0) = 13.7
\cdot  10^{-4}$ fm$^3$, and we used a phenomenological $\Delta\to N \gamma^*$ 
transition form factor $G_\Delta(Q^2)$ extracted from pion electroproduction in
the $\Delta$-resonance region. In the case of the proton's $\Sigma_p(Q^2)$ we
could compare with experimental values and found good agreement within the
accuracy of the data. The spin polarizabilities $\gamma_{p,n}(Q^2)$ pass
through zero (starting from  negative values) at $Q^2 = 0.4 \dots 0.5$ GeV$^2$ 
as a result of the cancelation between pion-loop and $\Delta(1232)$-effects. 
Furthermore, we found that there is a smooth transition from the hadronic side
to the partonic side at $Q^2 = 0.5 \dots 0.7$ GeV$^2$ for both $\Sigma_{p,n}
(Q^2)$ and $\gamma_{p,n}(Q^2)$. It means that these quantities are not very
sensitive to details of the nucleon excitation spectrum.  

\newpage
\section*{Appendix}
Here we collect explicit formulas for the functions $A_1(s,Q^2)$ and
$A_2(s,Q^2)$. The 52 one-pion loop diagrams evaluated in relativistic chiral
perturbation theory give rise to the following contributions for the proton, 

\begin{eqnarray} A_{2,p}(s,Q^2)&=& g_{\pi N}^2\bigg\{-\underline J_0(s)+{3\over
2} \underline J_1(s)+4\underline \gamma_3(s,Q^2)-2\underline \Gamma_3(s,Q^2)
+{1\over 2} (3M^2-Q^2-s) \Gamma_2(s,Q^2) \nonumber \\ & & +  (s-M^2+Q^2)\Big[
\Gamma_1(s,Q^2) -{Q^2\over 2} G_4(s,Q^2)+ {M^2\over 2} G_6(s,Q^2)\Big]  + Q^2
\Big[ \Gamma_5(s,Q^2) \nonumber \\ & & -2M^2 G_5(s,Q^2)\Big] + {Q^2\over s-M^2}
\Big[ {3\over 2} \underline J_1(s)+4\underline \gamma_3(s,Q^2)+2\underline 
\Gamma_3(s,Q^2)+M^2\underline \Gamma_6(s,Q^2) \nonumber \\ & & + Q^2\Big(
\underline \Gamma_5(s,Q^2)-\underline \Gamma_4(s,Q^2)\Big) \Big] + {3Q^2M^2 
\over (s-M^2)^2 } \Big[ \underline{\underline J}_1(s)- \underline{
\underline J}_0(s) \Big] \bigg\}\,\,, \\
\bigskip
A_{1,p}(s,Q^2)&=& A_{2,p}(s,Q^2) + g_{\pi N}^2 \Big\{(M^2-Q^2+
m_\pi^2-s) \big[ \Gamma_1(s,Q^2)-\Gamma_2(s,Q^2)\big] \nonumber \\ & & + 
2\underline \Gamma_3(s,Q^2)-4\underline \gamma_3(s,Q^2) +2m_\pi^2 \big[ 
\gamma_2(s,Q^2)-\gamma_1(s,Q^2)-\gamma_0(s,Q^2) \big] \Big\} \,\,, 
\end{eqnarray} 
and the neutron,
\begin{eqnarray} A_{2,n}(s,Q^2) &=& g_{\pi N}^2\Big\{4\underline\gamma_3(s,Q^2)
+4\underline \Gamma_3(s,Q^2)+ (3M^2-Q^2-s) \Gamma_2(s,Q^2) -4M^2Q^2 G_5(s,Q^2) 
\nonumber \\ & & + (s-M^2+Q^2)\big[2 \Gamma_1(s,Q^2) -Q^2 G_4(s,Q^2)+ M^2
G_6(s,Q^2)\big] \Big\} \,\,, \\ 
\bigskip
A_{1,n}(s,Q^2)&=& A_{2,n}(s,Q^2) + g_{\pi N}^2 \bigg\{2(M^2-Q^2+ m_\pi^2
-s) \big[ \Gamma_1(s,Q^2)-\Gamma_2(s,Q^2)\big] \nonumber \\ & & +
{8m_\pi^2 \over s-M^2+Q^2} \Big[ \underline \gamma_3(s,Q^2)- \underline 
\gamma_3(M^2-Q^2,Q^2)+ \underline \Gamma_3(s,Q^2)-\underline \Gamma_3(M^2-Q^2,
Q^2) \Big] \nonumber \\ & & -4\underline \Gamma_3(s,Q^2)-4\underline
\gamma_3(s,Q^2) +2m_\pi^2 \big[ \gamma_2(s,Q^2)-\gamma_1(s,Q^2)-\gamma_0(s,Q^2)
\big] \bigg\}  \,\,. \end{eqnarray} 
The various loop functions are generalizations of the ones defined for $Q^2=0$ 
in ref.\cite{pola} to virtual photons with $Q^2>0$ using $h_\gamma(x,y;s,Q^2) =
m_\pi^2(1-y) +M^2y^2+(s-M^2)xy(y-1) +Q^2x(1-x)(1-y)^2$ in their Feynman
parameter representation. The underbar on a loop function means
subtraction at $s=M^2$, e.g. $\underline \gamma_3(s,Q^2)=
\gamma_3(s,Q^2)-\gamma_3(M^2,Q^2)$. Similarly, the double underbar means two
subtractions at $s=M^2$, e.g. $\underline {\underline J}_0(s) = \underline
J_0(s) - (s-M^2) \underline J_0'(M^2)$. 

The tree diagrams with $\Delta(1232)$-resonance excitation when evaluated
relativistically using a Rarita-Schwinger spinor for the spin-3/2 field
give an equal contribution for the proton and the neutron,
\begin{eqnarray}
A_2(s,Q^2)&=&{\kappa^{*2}G^2_\Delta(Q^2)\over72M^2M_\Delta^2}
\bigg\{ {1\over s-M_\Delta^2} \Big[(s-M^2)^2(3M_\Delta^2-M^2) +Q^2(s M^2-3M^4 
\nonumber \\ & & -3 M^2Q^2  +3M^2M^2_\Delta+4sM_\Delta M -s Q^2+3s M_\Delta^2-
Q^4) \Big] -2(4Y+1) Q^2 \nonumber \\ & & \cdot (M^2+s+Q^2) -(4 Y+1)^2\big[Q^2
(s+M^2+4 M_{\Delta}M)+(s-M^2)^2\big] \bigg\}\,\,, \\
A_1(s,Q^2) &=& A_2(s,Q^2) -{\kappa^{*2}
G^2_\Delta (Q^2)\over 12M^2 (s-M_\Delta^2 )} \Big\{ Q^2(s+M^2+2M_\Delta M
)+(s-M^2)^2 \Big\}  \,\,. \end{eqnarray}

\newpage
\begin{figure}
\unitlength 1mm
\begin{picture}(160,55)
\put(0,0){\makebox{\psfig{file=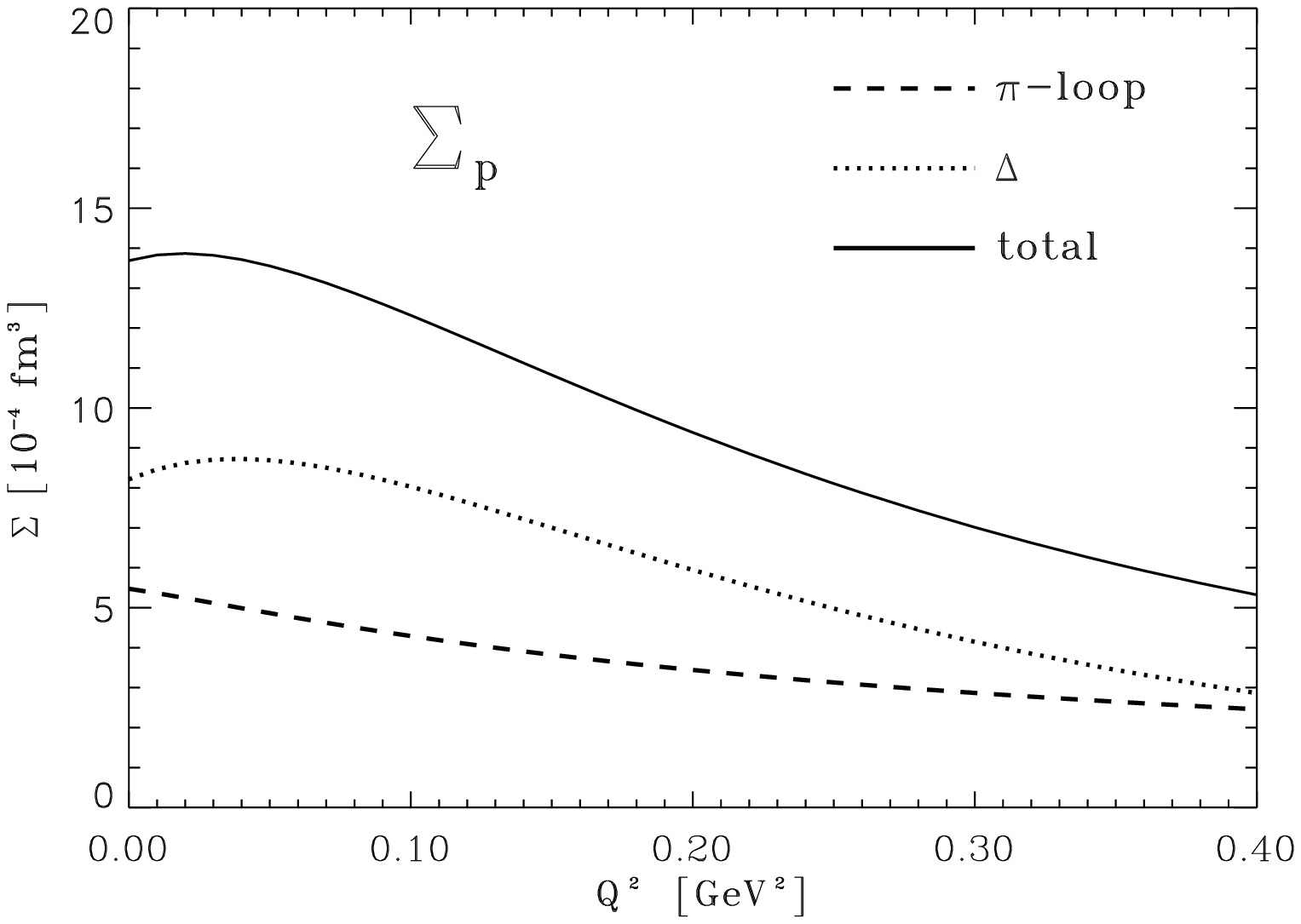,width=80.0mm}}}
\put(80,0){\makebox{\psfig{file=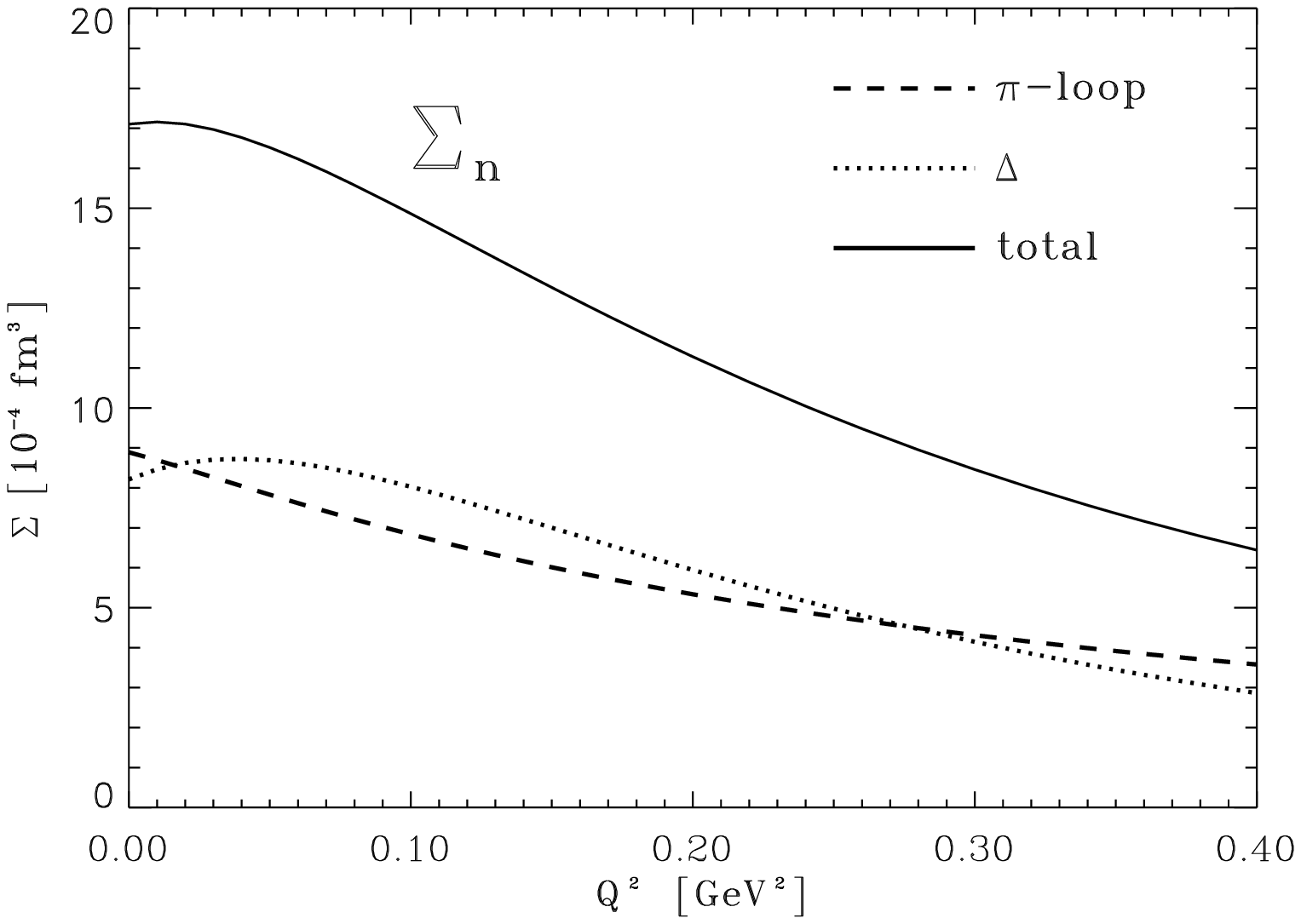,width=80.0mm}}}
\end{picture}

\medskip
{\it Fig.1: The sums of electric and magnetic polarizabilities
$\Sigma_{p,n}(Q^2)$ for the proton and the neutron as a function of $Q^2$. The
dashed, dotted and full lines are explained in the figure.}
\end{figure}

\bigskip

\begin{figure}
\unitlength 1mm
\begin{picture}(160,55)
\put(40,0){\makebox{\psfig{file=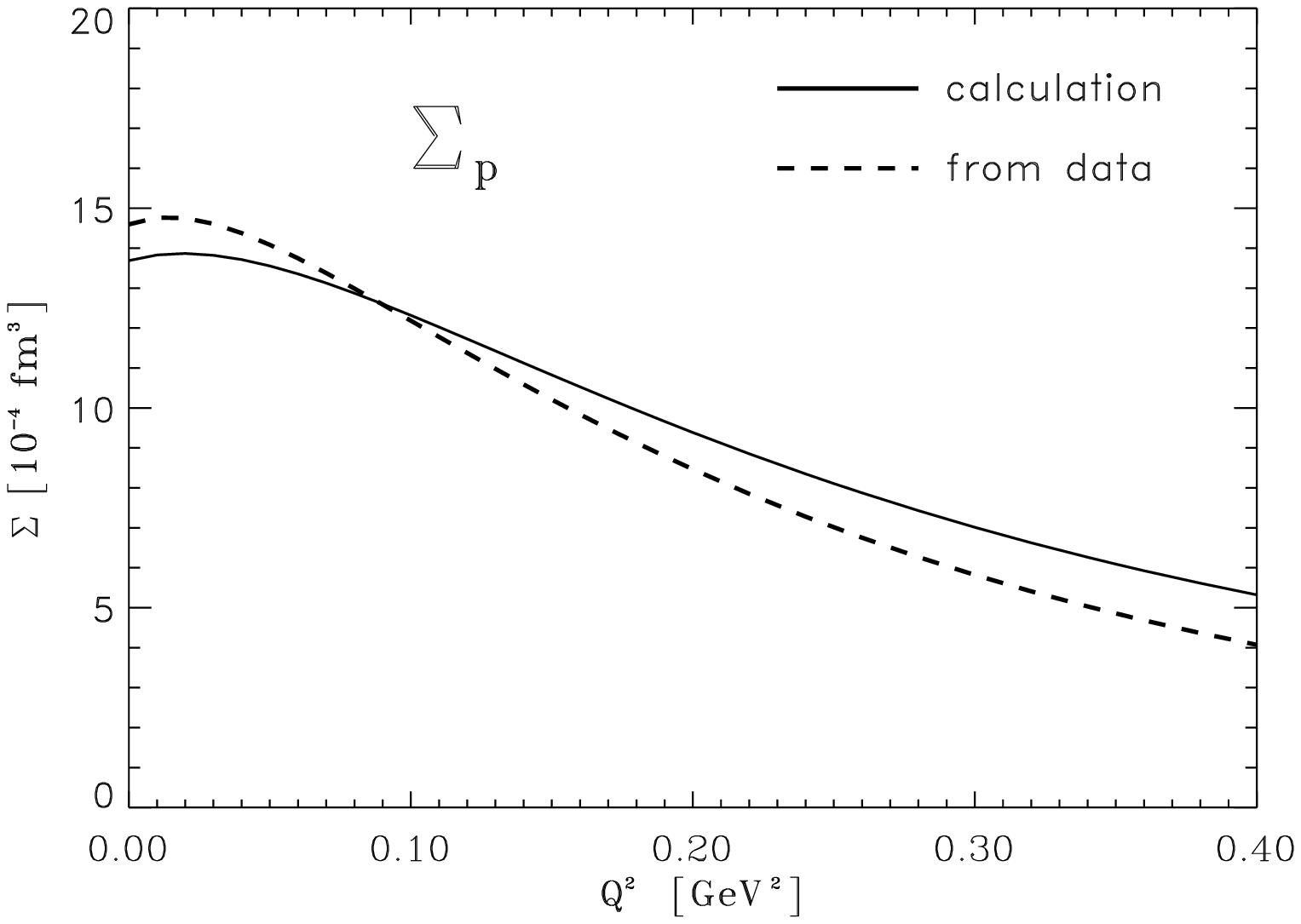,width=80.0mm}}}
\end{picture}
\medskip
{\it Fig.2: The calculated quantity $\Sigma_p(Q^2)$ (full line) compared to an
empirical determination using the electroproduction cross sections of
ref.\cite{data} (dashed line).} 
\end{figure}

\bigskip

\begin{figure}
\unitlength 1mm
\begin{picture}(160,55)
\put(0,0){\makebox{\psfig{file= 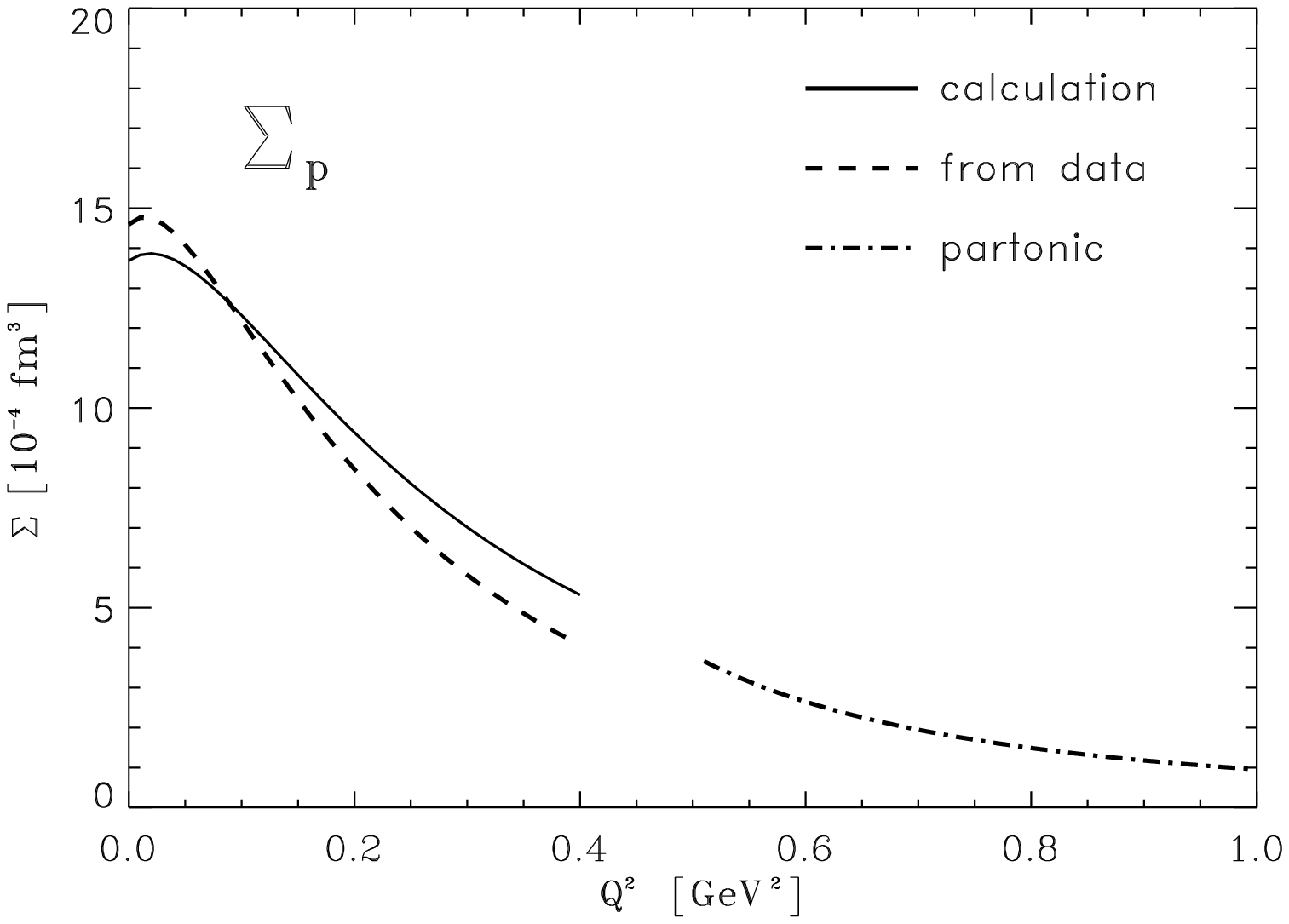,width=80.0mm}}}
\put(80,0){\makebox{\psfig{file=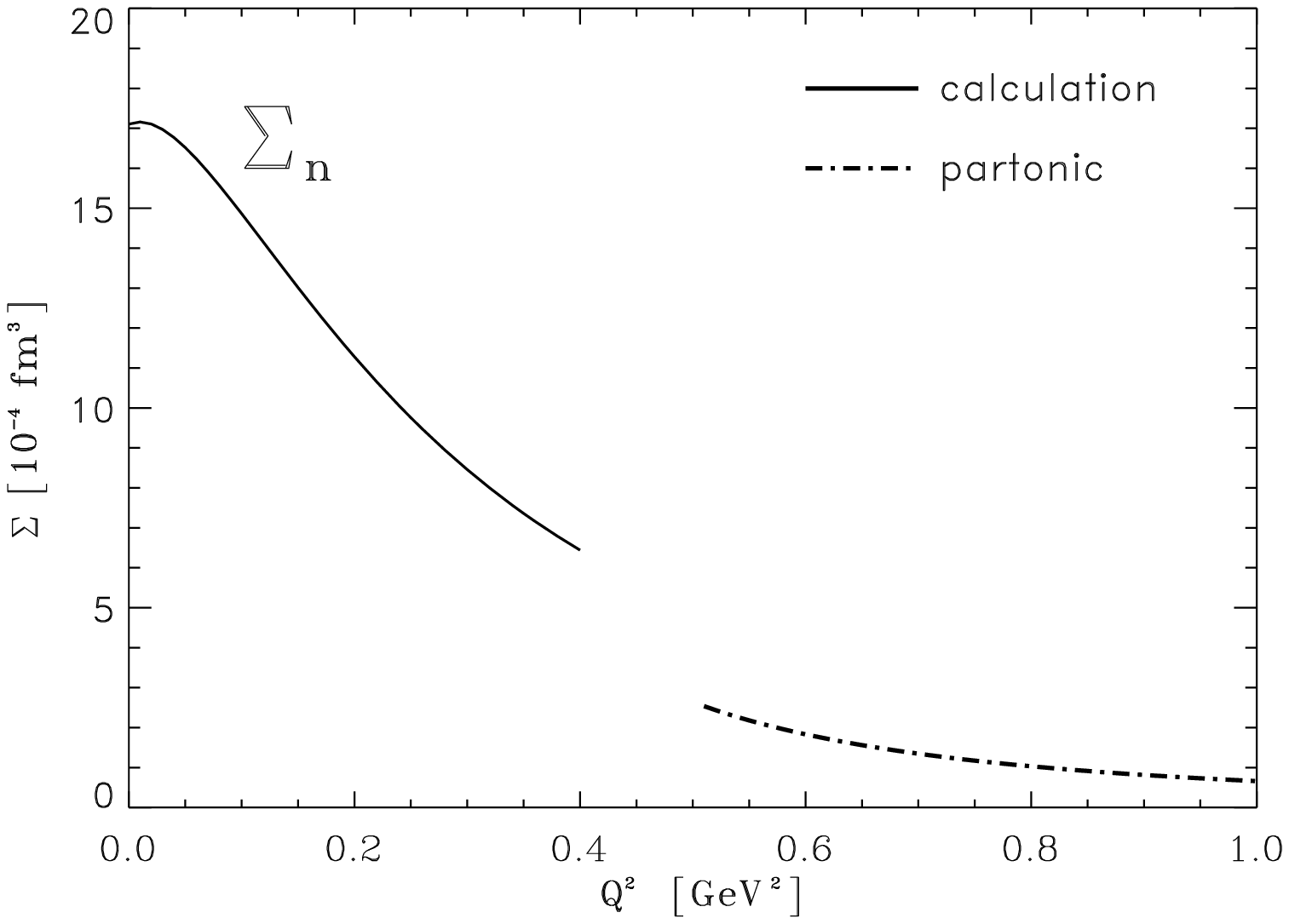,width=80.0mm}}}
\end{picture}

\medskip
{\it Fig.3: $\Sigma_{p,n}(Q^2)$ for the proton and the neutron as a function 
of $Q^2$. The prediction at small $Q^2$ (full lines) is due to relativistic 
pion loops and the $\Delta(1232)$-resonance. The dashed-dotted lines result
from extrapolating downward the partonic prediction.}
\end{figure}

\bigskip

\begin{figure}
\unitlength 1mm
\begin{picture}(160,55)
\put(0,0){\makebox{\psfig{file=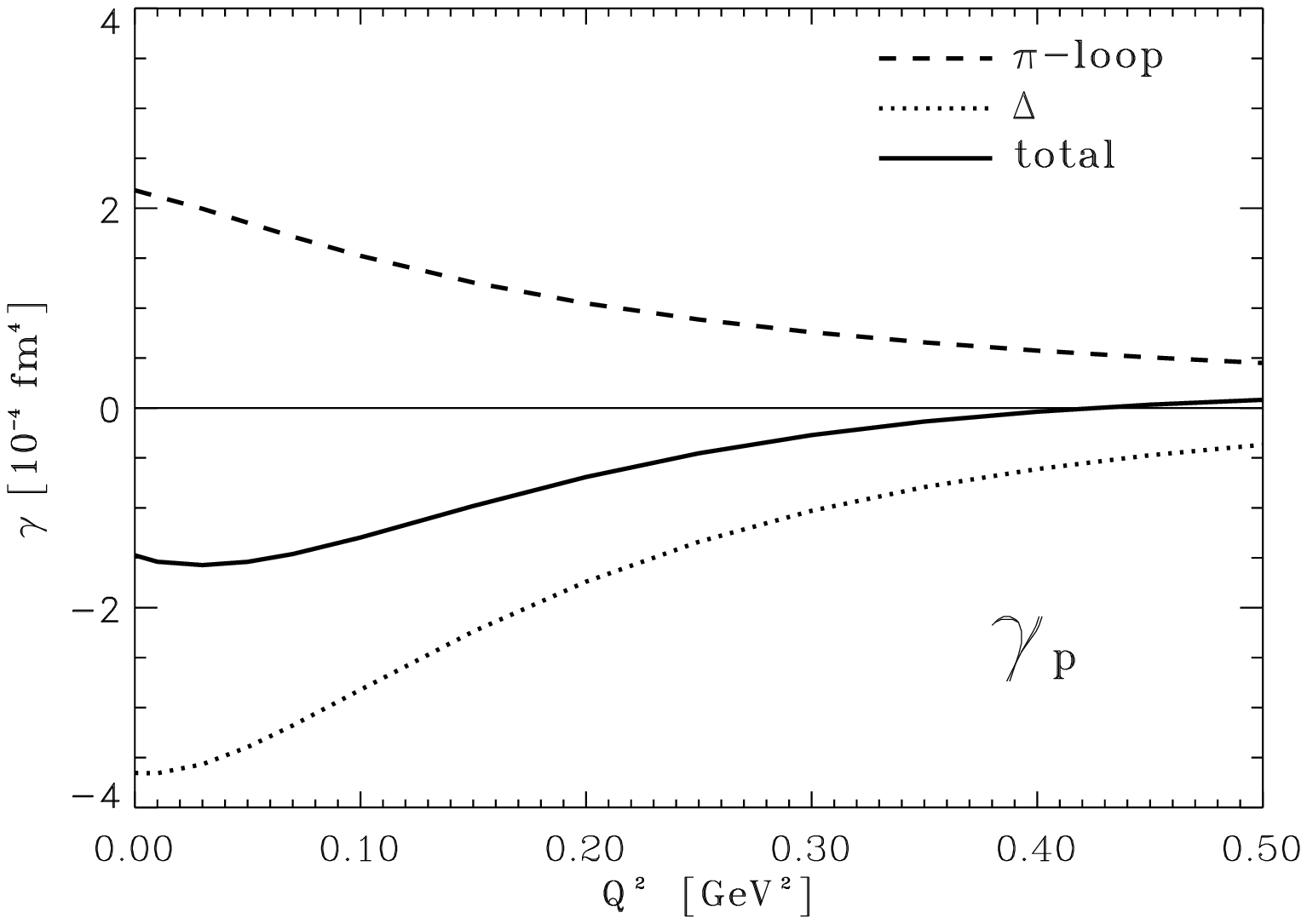,width=80.0mm}}}
\put(80,0){\makebox{\psfig{file=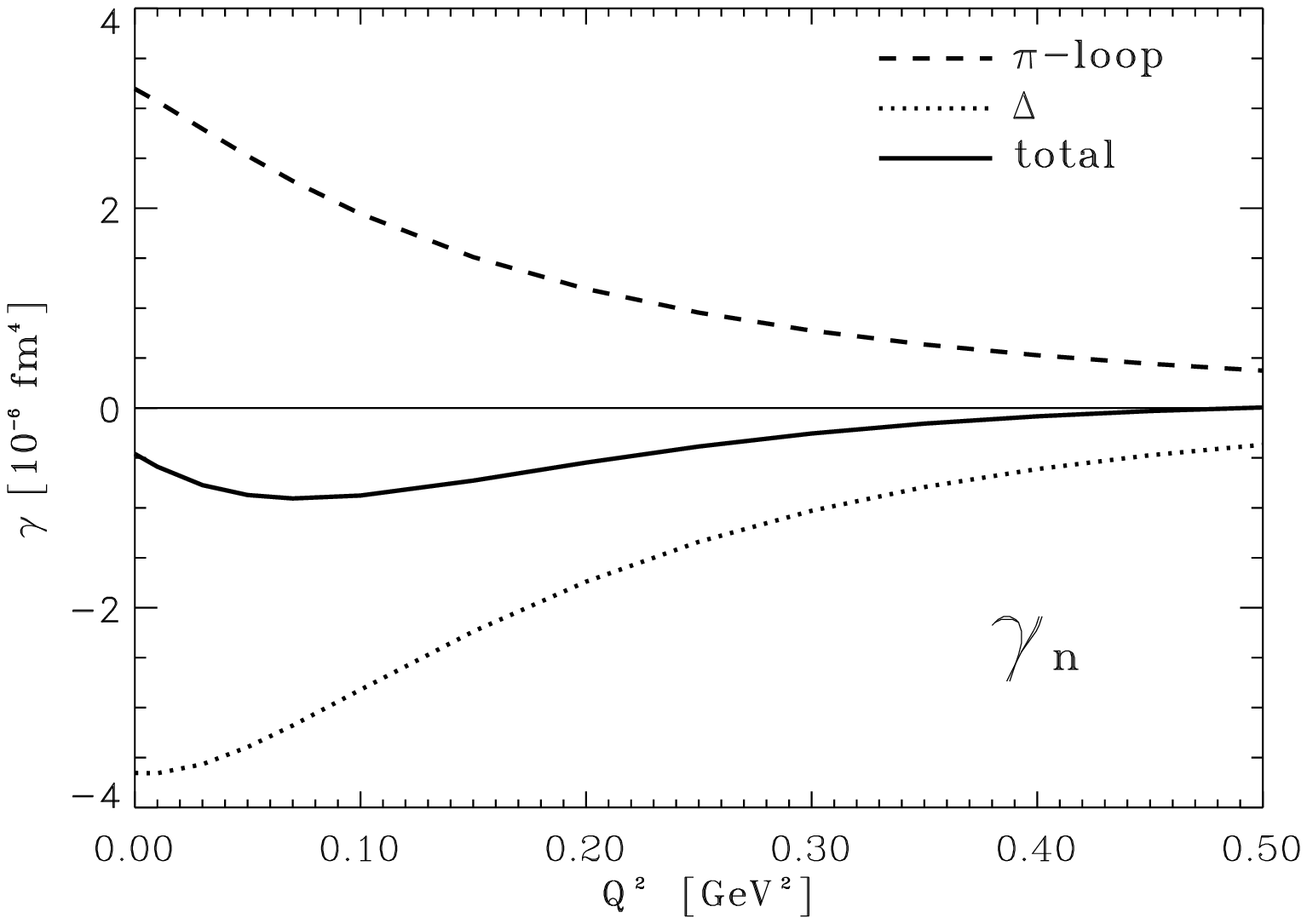,width=80.0mm}}}
\end{picture}

\medskip
{\it Fig.4: The generalized spin polarizabilities $\gamma_{p,n}(Q^2)$ for the
proton and the neutron as a function of $Q^2$. The dashed, dotted and full
lines are explained in the figure.} 
\end{figure}

\bigskip

\begin{figure}
\unitlength 1mm
\begin{picture}(160,55)
\put(0,0){\makebox{\psfig{file=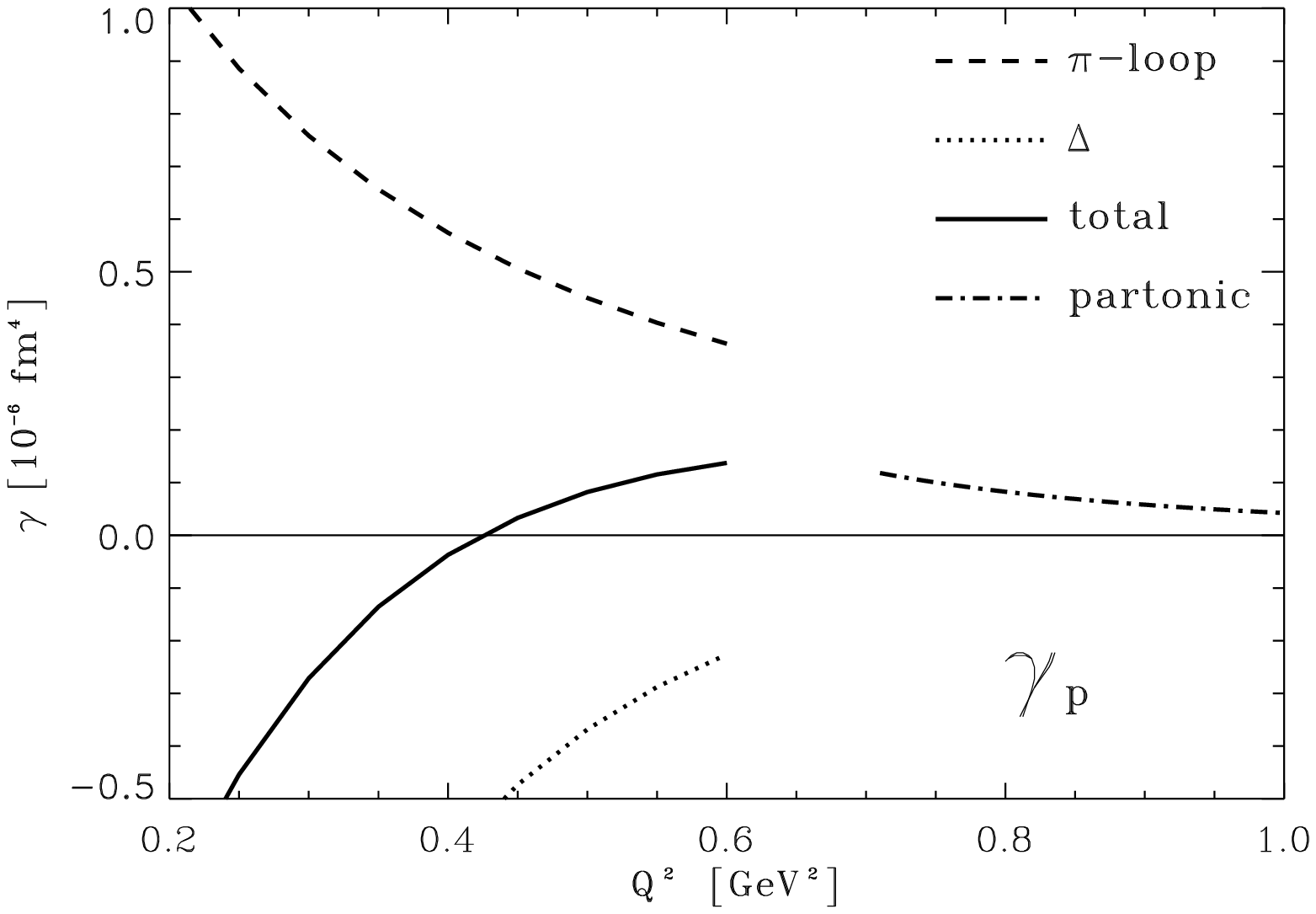,width=80.0mm}}}
\put(80,0){\makebox{\psfig{file=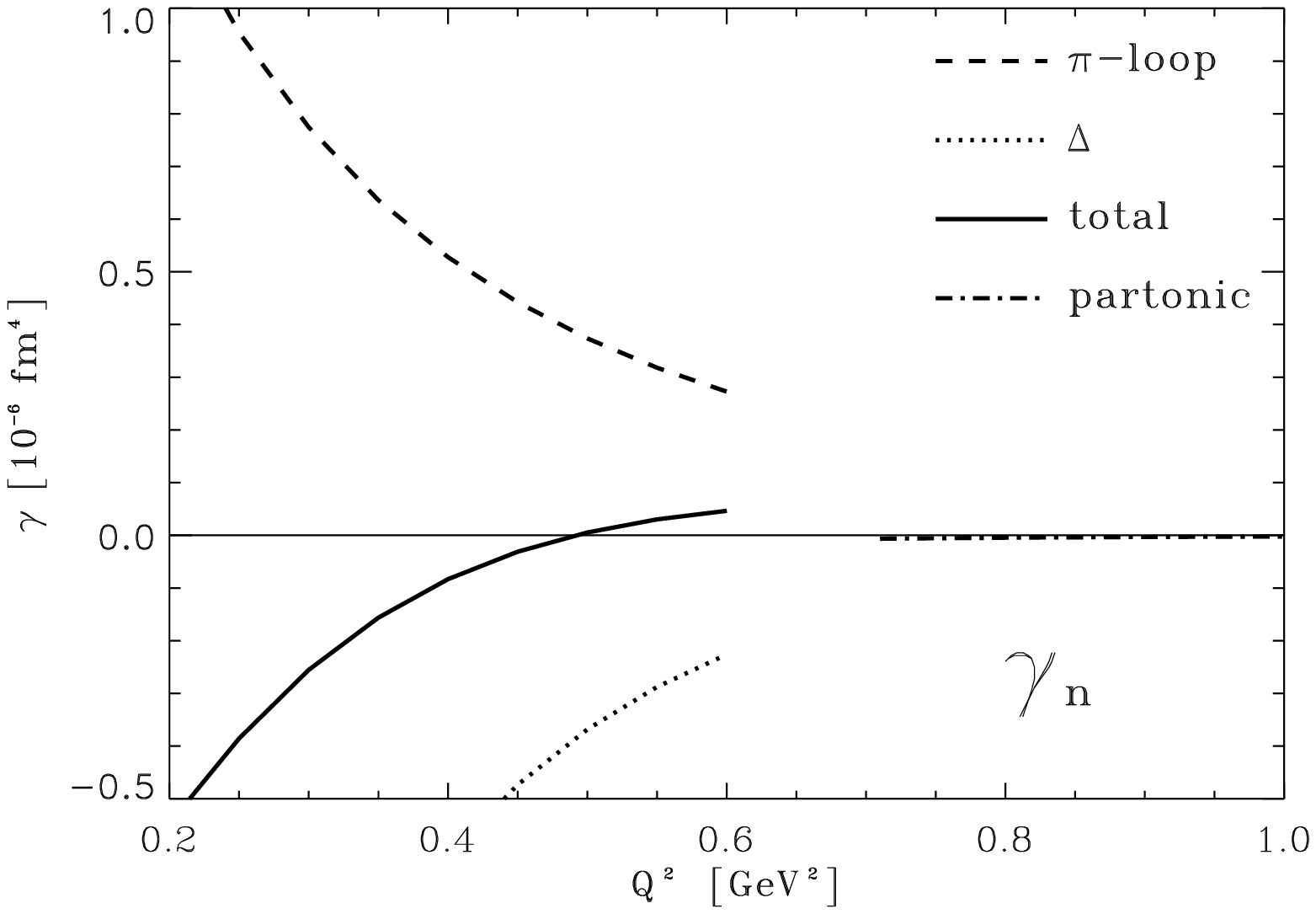,width=80.0mm}}}
\end{picture}

\medskip
{\it Fig.5: $\gamma_{p,n}(Q^2)$ for the proton and the neutron as a function 
of $Q^2$. The prediction at small $Q^2$ (full lines) is due to relativistic 
pion loops and the $\Delta(1232)$-resonance. The dashed-dotted lines result
from extrapolating downward the partonic prediction.}
\end{figure}

\end{document}